\documentclass[]{spie}  

\usepackage{amsmath,amsfonts,amssymb}
\usepackage{graphicx}
\usepackage[colorlinks=true, allcolors=blue]{hyperref}
\usepackage{wrapfig}

\newcommand{\lsim}{\mathrel{\hbox{\rlap{\lower.75ex \hbox{$\sim$}} \kern-.3em \raise.4ex \hbox{$<$}}}}
\newcommand{\gsim}{\mathrel{\hbox{\rlap{\lower.75ex \hbox{$\sim$}} \kern-.3em \raise.4ex \hbox{$>$}}}}

\title{VTXO: The Virtual Telescope for X-ray Observations}

\author[a,b,c]{John F. Krizmanic}
\author[c]{Neerav Shah}
\author[c]{Philip C. Calhoun}
\author[c]{Alice K. Harding}
\author[c]{Lloyd R. Purves}
\author[c]{Cassandra M. Webster}
\author[d,b,c]{Michael F. Corcoran}
\author[d,b,c]{Chris R. Shrader}
\author[e]{Steven J. Stochaj}
\author[e]{Kyle A. Rankin}
\author[e]{Daniel T. Smith}
\author[e]{Hyeongjun Park}
\author[e]{Laura E. Boucheron}
\author[e]{Krishna Kota}
\author[f]{Asal Naseri}
\affil[a]{University of Maryland,  Baltimore County, Baltimore, MD 21250 USA}
\affil[b]{Center for Space Science \& Technology (CRESST)}
\affil[c]{NASA/Goddard Space Flight Center,Greenbelt MD 20771 USA}
\affil[d]{The Catholic University of America, Washington, DC 20064 USA}
\affil[e]{New Mexico State University, Las Cruces NM 88003 USA}
\affil[f]{Space Dynamics Laboratory Logan UT 84341 USA}

\authorinfo{Further author information: (Send correspondence to John F. Krizmanic)\\John F. Krizmanic: E-mail: john.f.krizmanic@nasa.gov, Telephone: 1 301 286 6817}

\begin{document} 
\maketitle

\begin{abstract}
The Virtual Telescope for X-ray Observations (VTXO) will use lightweight Phase Fresnel
Lenses (PFLs) in a virtual X-ray telescope with $\sim$1 km focal length and with $\sim$50 milli-arcsecond
angular resolution. VTXO is formed by using precision formation flying of two SmallSats: a
smaller OpticsSat that houses the PFLs and navigation beacons while a larger DetectorSat
contains an X-ray camera, a precision start tracker, and the propulsion for the formation flying.
The baseline flight dynamics uses a highly elliptical supersynchronous orbit allow the formation
to hold in an inertial frame around the 90,000 km apogee for 10 hours of the 32.5 hour orbit with
nearly a year mission lifetime. VTXO's fine angular resolution enables measuring the
environments close to the central engines of bright compact X-ray sources. This X-ray imaging
capability allows for the study of the effects of dust scattering near to the central objects such as
Cyg X-3 and GX 5-1, for the search for jet structure near to the compact object in X-ray novae
such as Cyg X-1 and GRS 1915+105, and for the search for structure in the termination shock of
in the Crab pulsar wind nebula.  The VTXO SmallSat and instrument designs, mission
parameters, and science performance are described. VTXO development was supported as
one of the selected 2018 NASA Astrophysics SmallSat Study (AS$^3$) missions.  
\end{abstract}


\section{VTXO Science enabled by Milli-arcsecond Angular Resolution}
\label{sec:Science}  

\begin{figure}
\begin{center}
   \includegraphics[width=0.95\columnwidth]{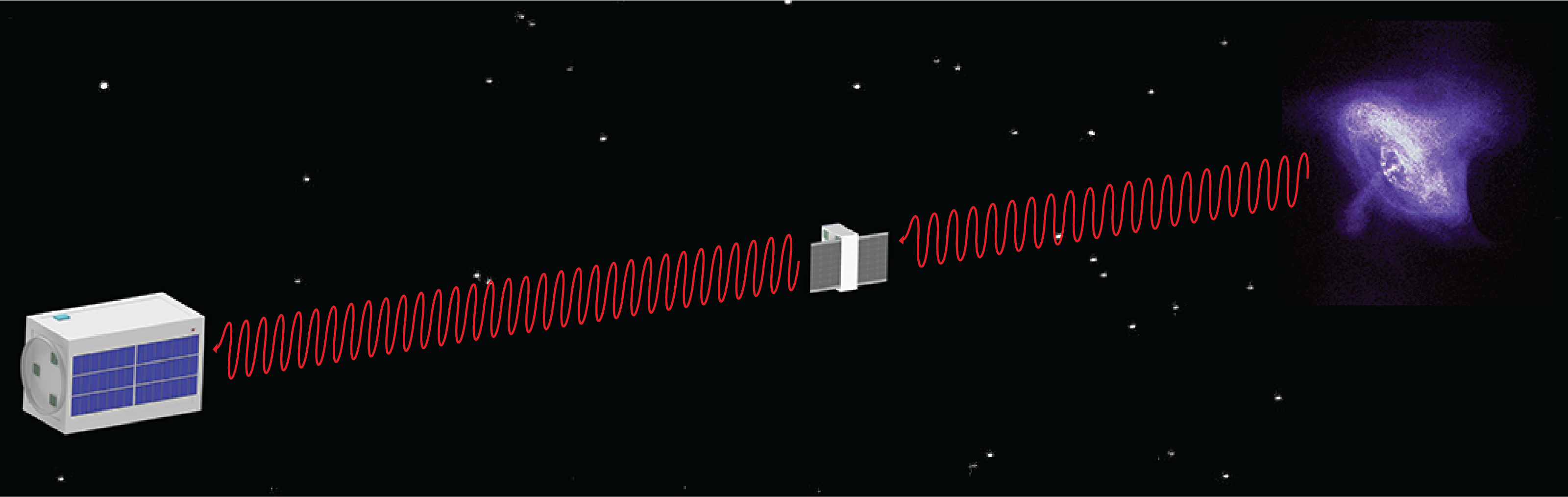}
\end{center}
\caption{CAD rendition of the two VTXO SmallSats flying in formation viewing the Crab nebula. The DetectorSat is behind the OpticsSat forming a virtual X-ray telescope with 1 km focal length (not to scale). From Ref. \citenum{2020arXiv200612174K} .}
\label{CADformation}
\end{figure}

The goal of VTXO is to employ precision  formation-flying SmallSats to form an X-ray telescope that can perform high resolution imaging of some of the brighter X-ray sources to significantly increase our understanding of the underlying astrophysics. Current X-ray telescopes using conventional Wolter type-1 X-ray optics, i.e. ``grazing incidence'' optics, such as those employed in the EINSTEIN X-ray telescope \cite{Giacconi1980}, Chandra X-ray Observatory \cite{Swartz2010}, and proposed Lynx X-ray Observatory \cite{2019JATIS...5b1001G} limit the angular resolution in the X-ray band to $\gsim 0.5$ milli-arcseconds (mas). Development of ground-testable PFLs and laboratory characterization measurements have demonstrated near diffraction-limited angular resolution in the X-ray band \cite{2020arXiv200812810K}.  However, PFLs require long focal lengths and those sized for VTXO lead to a 1 km focal length requiring precision formation flying of two SmallSats to form the X-ray telescope.  VTXO employs a 6U OpticsSat that houses the PFLs and an ESPA-class DetectorSat that contains an X-ray camera, a precision star tracker, and propulsion for the formation flying. The baseline flight dynamics is based on a highly elliptical orbit, with 90,000 km apogee and 600 km perigee altitudes, which allows the formation to hold in an inertial frame for $\pm 5$ hours, e.g. 10-hour interval, around apogee during each 32.5 hour orbit.  Figure \ref{CADformation} shows a CAD rendition of the two VTXO SmallSats flying in formation.

VTXO 50-mas angular resolution is an order of magnitude improvement over that of the current state of the art, e.g. that provided by the Chandra Observatory. This ability to probe at $\times 10$ finer spatial scales provides measurements of the high energy environments closer to the central engines of compact X-ray sources.  VTXO's fine imaging capability will be first proofed on orbit via imaging Sco X-1 when the SmallSats are in formation.  Subsequent astronomical measurements include a number of objects that VTXO can reasonably image on the 10-hour time scale, and these are detailed in Table~\ref{SourceTable} in consideration of the energy bandpass and efficiencies of the PFL optics.  VTXO targets include Cygnus X-1 to survey the wind environment, the Crab pulsar to search for sub 0.1 arcsecond structures in the plerion nebula, Cygnus X-3, GX 5-1 \& Centaurus X-3 to study of the effects of dust scattering halos nearer to these compact objects, Eta Carinae to image the of bow shocks in this colliding wind binary, and to target transients such as V404 Cygni to search for scattering echoes and jet structures nearer to the central object, More challenging targets that are thought to require better angular resolution and sensitivity include Gamma Cassiopeiae to resolve structure in the enigmatic Be star,  and space weather studies on nearby exoplanets such as Proxima B. 

\begin{table}[h]
\begin{center}
{\bf \Large
\begin{tabular}{|l|c|c|} \hline
\multicolumn{3}{|c|}{\bf \Large Observation time for 1000 VTXO counts} \\
\multicolumn{3}{|c|}{\bf \Large in the energy band 4.5 +/- 0.075 keV} \\ \hline
Source &  Flux (mCrabs) & Obs Time (hr)  \\ \hline 
Sco X-1  & 8000 & 0.2 \\ \hline
GX 5-1 & 1260 & 1.5 \\ \hline
GRS 1915+105 & 450 & 4.2 \\ \hline
Cyg X-3 &  390 & 4.9 \\ \hline
Cyg X-1  & 350 & 5.4 \\ \hline
Crab Pulsar & 100 & 19 \\ \hline
Cen X-3 & 90 & 21 \\ \hline
$\gamma$Cas & 13 & 146 \\ \hline
Eta Carinae & 4.2 & 452 \\ \hline
\end{tabular}
}
\end{center}
\caption{Baseline ability of VTXO to observe 1000 counts from bright compact X-ray sources assuming a 150 eV FWHM energy resolution around 4.5 keV and a 3 cm diameter PFL with 30\% efficiency. 1 crab is defined to be $2.4 \times 10^{-8}$ ergs/cm$^2$/s over the energy range 2 to 10 keV.} 
\label{SourceTable}
\end{table}

\subsection{PFL Optics Performance}

Phase Fresnel Lenses (PFLs) employ diffraction to focus incident
radiation to a primary focal point and
can be used as the optics of an X-ray or gamma-ray
telescope with potentially diffraction-limited angular resolution \cite{Skinner2001, Skinner2002,Krizmanic2005b}. However,  the requirement of relatively large PFL diameters to yield sufficient X-ray sensitivity yield long focal lengths for the telescope, leading to the requirement of formation=flying spacecraft.
The thickness profile of a PFL is varied as a function of
radius to form individual Fresnel zone in order to change the phase of the incident radiation to
form the primary focus. 
The focal length is defined by f = R p$_{\rm MIN}$/$\lambda$ where R is the
PFL radius, p$_{\rm MIN}$, is the pitch of the outermost Fresnel zone, and $\lambda$ is the wavelength to be
imaged. 
The diameter of the PFL is
determined by the minimum p$_{\rm MIN}$ that can be achieved with the fabrication technique, the energy
to be imaged, and the focal length.
The maximal height of the Fresnel ridges is material dependent, and the thickness for a
2$\pi$ phase change is given by t$_{2\pi}$ = $\lambda/\delta$ where $\delta$ is the index of refraction decrement of the material. For
example, a 3 cm diameter PFL designed to image at 4.5 keV with p$_{\rm MIN}$ = 20 microns would have
a 1.1 km focal length with t$_{2\pi} \approx 18$ microns assuming the lens material is polyimide. In principle, the
imaging efficiency can be nearly 100\% if the PFL radial profile is exact and absorption losses are negligible. In
practice, each Fresnel zone profile is made with steps, with the PFL efficiency dependent on the
number of steps used for each zone from 40\% (2 steps) to 95\% (8 steps). 

\begin{wrapfigure}{r}{0.65\textwidth}
\vspace{-4 mm}
\begin{center}
   \includegraphics[width=0.30\columnwidth,trim=0 0.5cm 0 1cm]{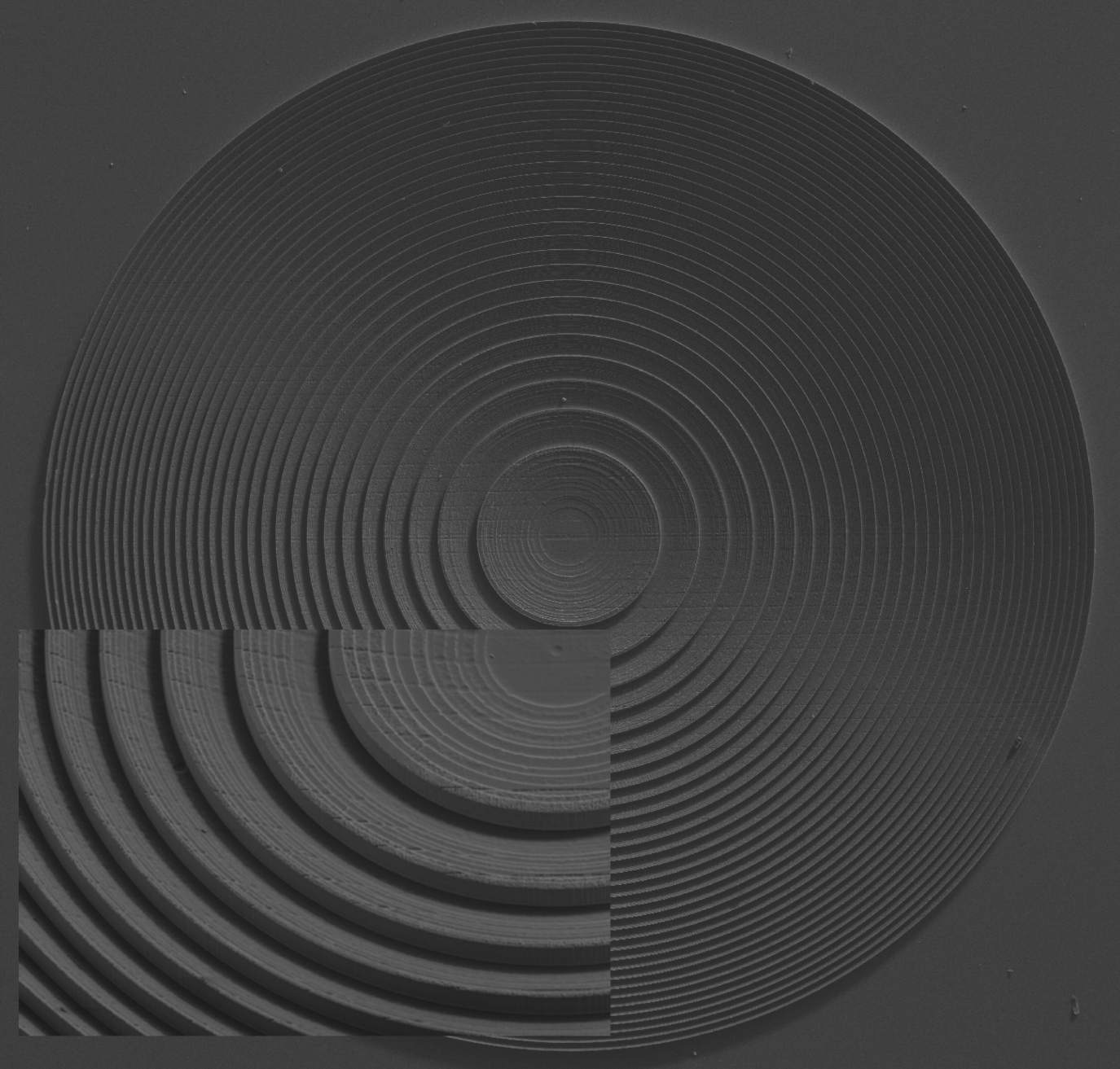}
    \includegraphics[width=0.30\columnwidth,trim=0 1.5cm 0 1cm]{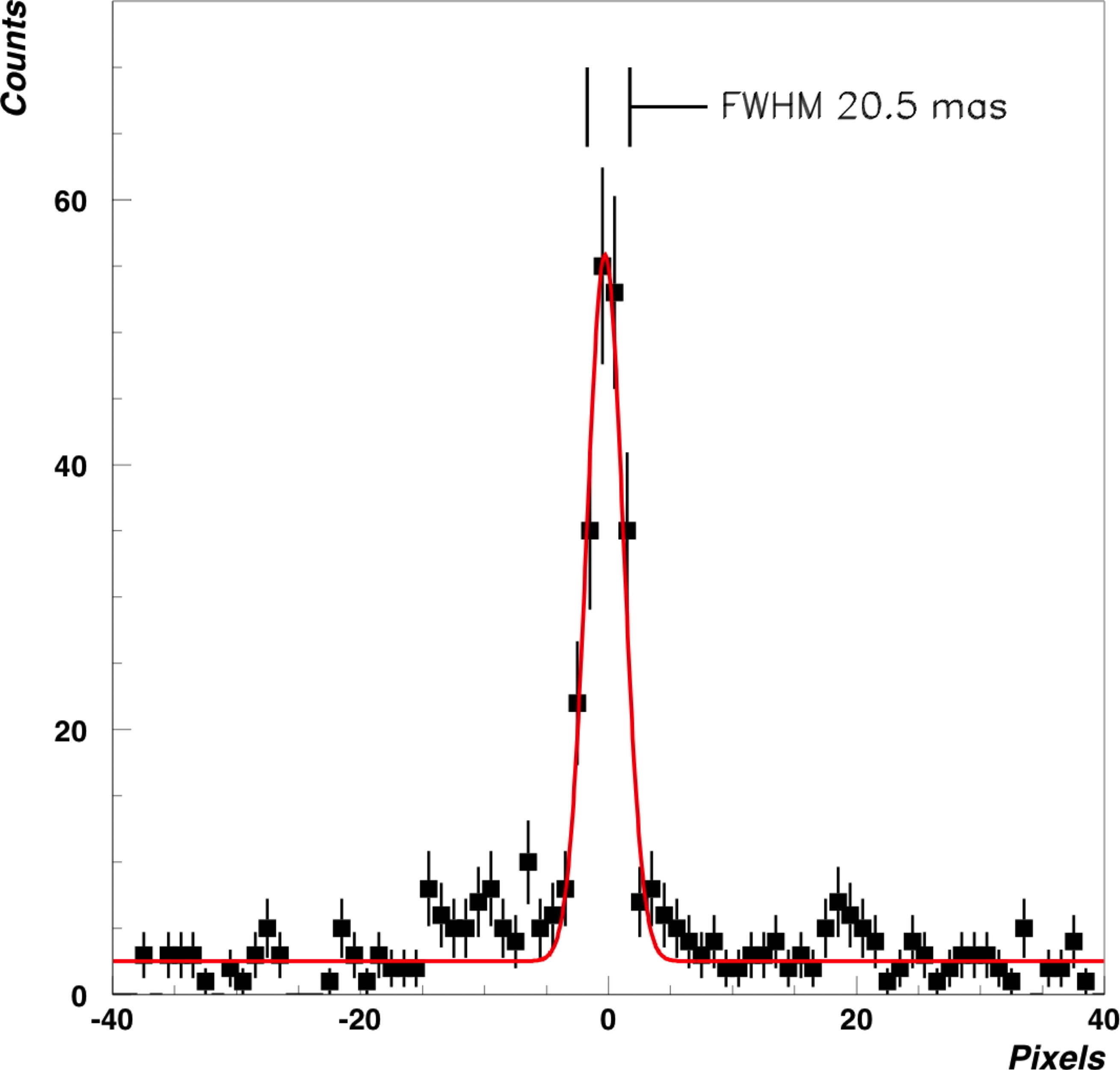}
\end{center}
\caption{Left, an SEM of a 3 mm diameter PFL designed to image at
8 keV, inset shows magnified part of the PFL; Right, the measured
point-spread-function of 20.5 milli-arcseconds (mas) which is near the
diffraction limit value of 16 mas. From Ref.~\citenum{2020arXiv200812810K}. }
\label{PFLimaging}
\vspace{-3 mm}
\end{wrapfigure}

We have fabricated X-ray imaging PFLs in silicon using MEMS fabrication techniques \cite{Morgan2004}
and
characterized their performance in a 600-m X-ray beam
line \cite{Krizmanic2009}, achieving near-diffraction limited imaging with high efficiency.
For this laboratory test, a 3 mm
diameter PFL had a focal length of
110.4 meters and was designed to
image at 8 keV (Cu K$\alpha$). Figure \ref{PFLimaging}
shows a SEM of the PFL along with
the measured imaging performance.
The efficiency of the PFL (and
substrate) was determined to be 36\%.
We have also fabricated and characterized a refractive/diffractive contact pair achromat and
have determined that the energy range that the PFL images can be increased to a 2 keV wide band
\cite{Krizmanic2009}.  
Subsequent PFL development at GSFC has fabricated and characterized a 3 cm diameter PFL designed for 4.5 keV and with $\sim$1 $\mu$m minimum feature size \cite{Dennis2012}.
For the baseline VTXO imaging performance, we assume a 3 cm diameter lens based on the 3 cm diameter and a conservative 30\% efficiency is assumed, which provide for the results presented in Table~\ref{SourceTable}. VTXO will employ three different PFL optics: the 4.5 keV design energy of the first PFL is chosen to optimize the count rate, assuming a power law X-ray spectrum, and the 6.7 keV design energy for the second PFL is chosen to optimize on the 6.7-keV emission line from helium-like iron. VTXO will also carry a third PFL-achromat optic to perform extended X-ray bandwidth measurements of the sources. These optics will allow for VTXO acquires 1000 counts in 1.9 hours in the band 4.5 $\pm$  0.075 keV and in 4.9 hours for the band 6.7 $\pm$ 0.075 keV for a 1 crab strength source and assuming 30\% imaging efficiency.While the narrow bandpass around the PFL imaging design energies limit the VTXO observations to brighter X-ray sources, the background due to the diffuse X-ray background is nearly non-existent, less than one event for a 10 hour observation. We plan to measure the diffuse X-ray background ($E \gsim 3$ keV) for a short duration exposure during each VTXO observation. The field of view (FoV) of VTXO is determined by the size of the X-ray camera , (DFP $\approx 4$ cm) for the single Teledyne H2RG HyViSI sensor, and the focal length (f$= 1$ km)) of the telescope leading to DFP /f = 8 arcseconds.

The angular resolution of a PFL in a telescope is determined by three terms \cite{Skinner2001,Skinner2002}: the diffraction limit of the PFL  ($\theta_{Diff} =1.22 \lambda/d$), the chromatic aberration of the PFL ($\theta_{CA}=0.2 \Delta (E/E) (d/f)$), and that due to the effects of a finite pixel size ($\theta_{Pxl} = \Delta x/f$),  $E$ ($\lambda$) is the PFL design  energy (wavelength) to be imaged, $d$ is the diameter of the PFL, $f$ is the focal length, $\Delta E$ is the energy resolution of the X-ray camera, and $\Delta x$ is the linear pixel size in the camera.

\begin{wrapfigure}{r}{0.5\textwidth}
\vspace{-10 mm}
\begin{center}
    \includegraphics[width=0.47\columnwidth]{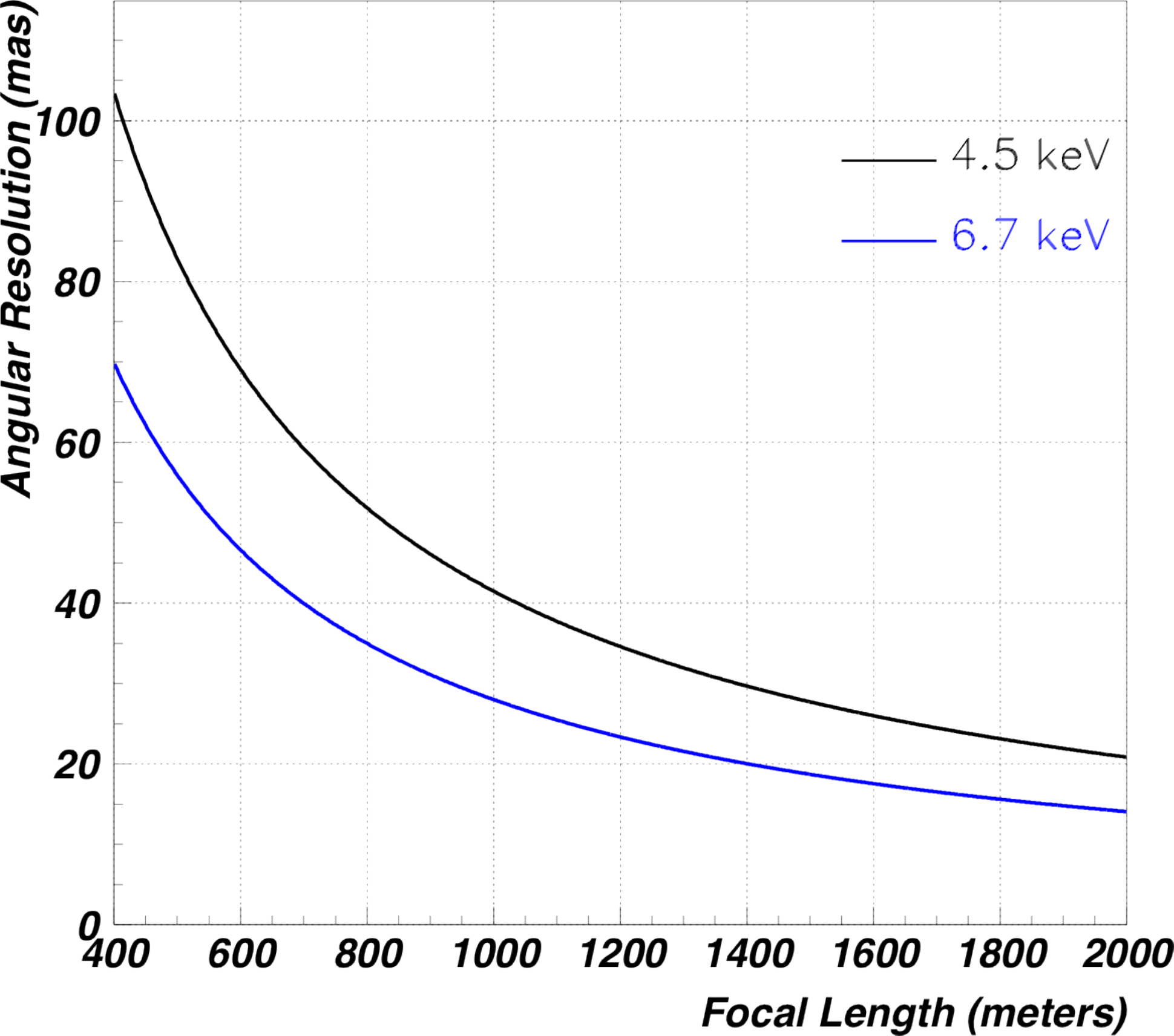}

\end{center}
\caption{The angular resolution of VTXO as a function of focal length at 4.5 keV and 6 keV assuming a 3 cm diameter PFL  and 150 eV energy resolution of the X-ray camera.From Ref. \citenum{2020arXiv200612174K}. }
\label{AngRes}
\end{wrapfigure}

The angular resolution obtained by combining the three terms in quadrature (and  and is dominated by the chromatic aberration) is shown in Figure \ref{AngRes} for the energies 4.5 keV and 6,7 keV as a function of focal length of the telescope. The results show the benefits of longer focal lengths and drives the choice for 1 km focal length, especially in consideration of that the the virtual telescope alignment errors dilute the PFL imaging performance yielding 55 mas as opposed to the 42 mas obtained from Figure \ref{AngRes}. The source of the  (formation-flying alignment errors will be discussed in the next section.  In this context, it should be noted that the PFL imaging performance is fairly insensitive to modest tilt angles, e.g. not imposing a strict requirement on controlling this orientation of the PFL and X-ray camera. The imaging error is also fairly insensitive to an error in the focal length since this effectively corresponds to an energy error and needs to be relatively small compared to the other terms, e.g. a 10 m distance error over 1 km corresponds to a 0.1\% energy error which is much smaller than that in the imaging error due to chromatic aberration.

\section{VTXO Instrumentation and SmallSat Spacecraft}
\label{sec:Science}  

\section{VTXO SmallSats and Instruments}

\begin{figure}
\begin{minipage}[t]{0.47\textwidth}
    \includegraphics[width=0.97\columnwidth]{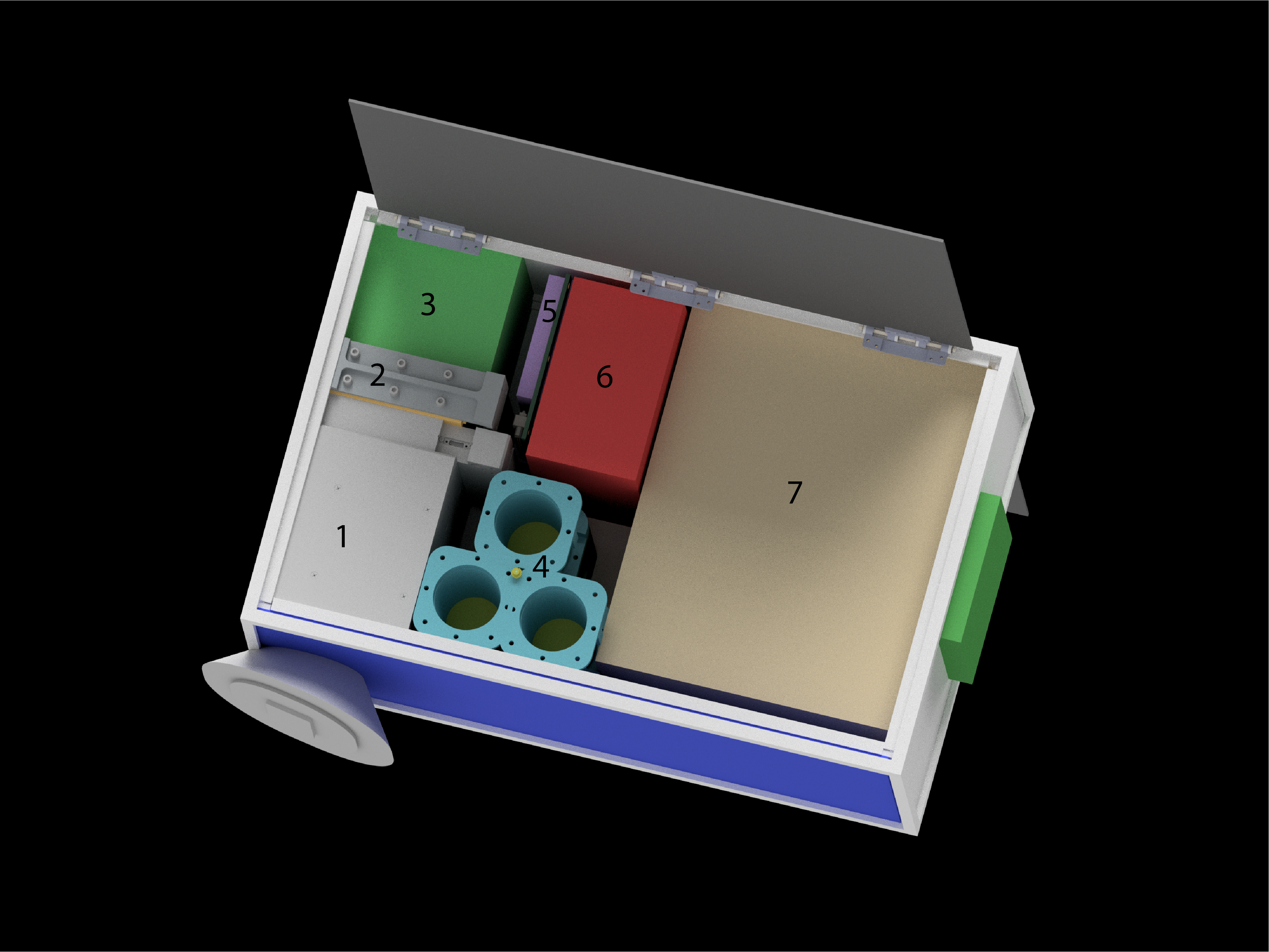}
\caption{\small {Internal view of the VTXO OpticsSat.  Key:BCT XACT-50  (star camera FoV shown by cone); 2 GPS receiver; 3 SWIFT SLX radio; 4 PFL assembly and laser beacons; 5 EPS unit; 6 Batteries; 7 VACCO cold gas MiPS. }}
\label{OptSatInt}
\end{minipage}
\hfill 
\begin{minipage}[t]{0.47\textwidth}
     \includegraphics[width=0.97\columnwidth]{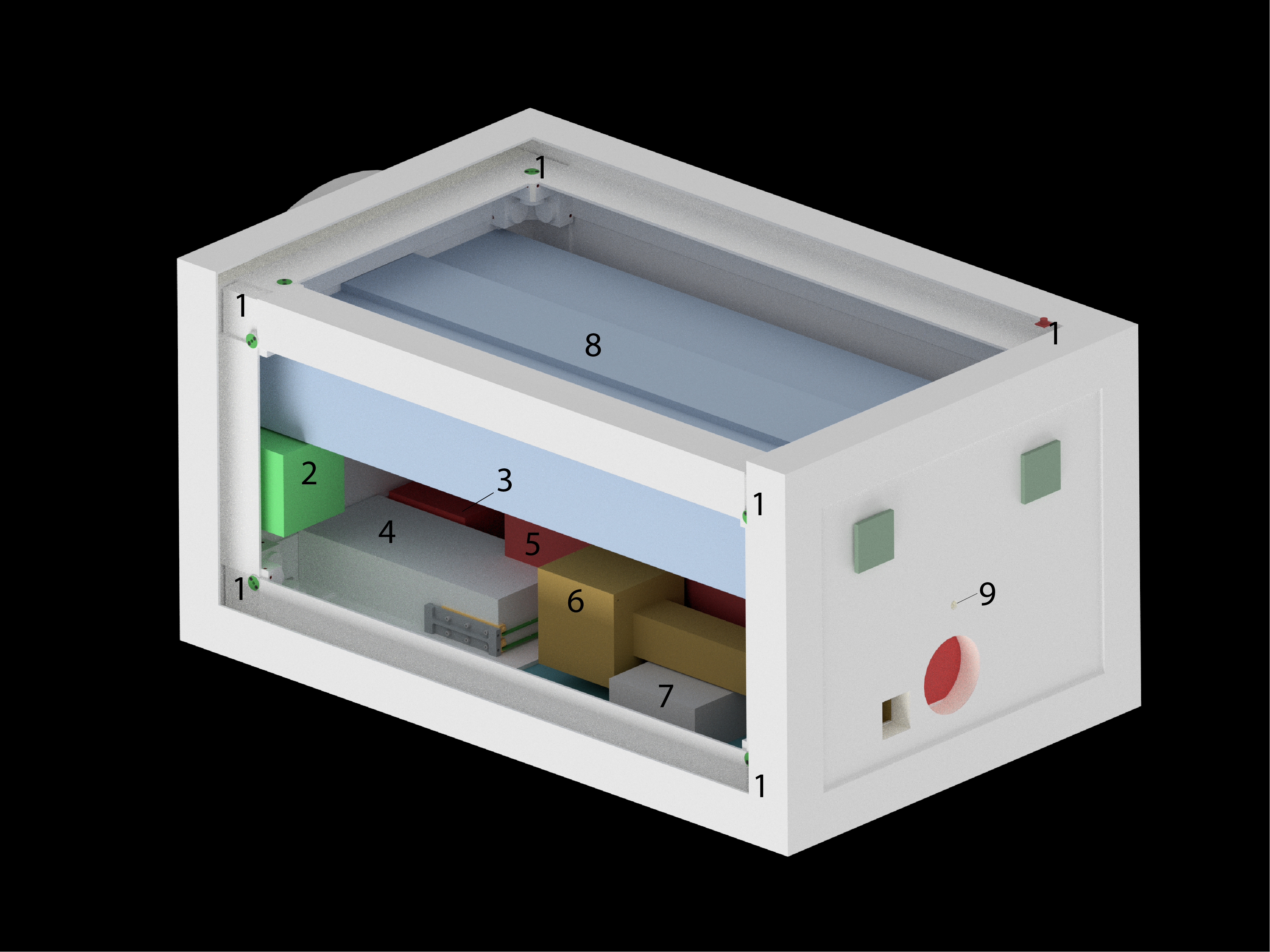}
\caption{\small {Internal view of the VTXO DetectorSat.  Key: 1 VACCO 3-nozzle cold gas thruster; 2 SWIFT SLX radio; 3 Batteries; 4 Avionics bus/shielding; 5 NISTEx-II interferometric start tracker; 6 X-ray camera assembly; 7 Instrument electronics/shielding; 8 VACCO cold gas generator; 9 X-ray camera/star tracker viewing ports. }}
\label{DetSatInt}
\end{minipage}
\end{figure}

The VTXO 6U OpticsSat, shown in Figure \ref{OptSatInt}, houses three 3-cm diameter PFL optics, laser beacons, avionics, thermal control, power, and propulsion systems. As previously discussed, the PFLs are designed for two specific energies, 4.5 keV and 6.7 keV, and achieve the 50-mas scale imaging in narrow 0.15 keV bands around the design energies. The third optics port will be used to characterize a PFL-based achromat that would extend the imaging performance over a $\sim$ 2 keV bandwidth. The VTXO ESPA-class DetectorSat, shown in Figure \ref{DetSatInt} hosts a finely pixelated X-ray camera using a Teledyne H2RG HyViSI sensor with ACADIA ASIC readout, a NISTEx-II precision star tracker, a star tracker based relative navigation system, thermal control, power, and a larger propulsion system used to perform the formation flying.  The NISTEx-II precision star tracker images the star field and laser beacons on the OpticsSat to achieve the virtual telescope alignment.  The NISTEx-II has a theoretical alignment performance of 41 mas and is currently operating on the STP-H6 platform on the International Space Station \cite{NISTEX}. The DetectorSat also includes a trapped particle radiation detector to assess the background from ambient environment around apogee  during science operations. It should be noted that the VTXO orbit was chosen with a 90,000 km altitude
apogee that spends more than 50\% of the orbit period outside the outer trapped electron radiation belt, Analysis of the anticipated radiation dose acquired in the baseline VTXO orbit has been determined to be $< 20$ kRad/year assuming at least 5 mm  of aluminum shielding.

\section{VTXO Mission and Observation Performance}

The mission concept of operations has the two VTXO SmallSats to be included on an ESPA ring on a ride-share launch that uses a supersynchronous GTO with a 90,000 altitude apogee and 200 km perigee, such as that obtained by SpaceX FH2 flight launched April 11, 2019. After deployment from the ESPA ring and satellite acquisition, the perigee will be raised to 600 km altitude for both satellites during the VTXO commissioning phase followed by commissioning of the formation flying and science instruments. Assuming 60 days for the commissioning, VTXO science operations will begin with an estimated science mission lifetime of $\sim$ 200 days based on a determination of the propulsion requirements, performance, and baseline fuel allocation. Ongoing trade  studies optimizing the formation flying GN\&C have shown the potential to increase the science mission lifetime by$\sim$50\% \cite{2020arXiv200709287R,2020arXiv200709289R}.  It should be noted that VTXO has the inherent capability to perform target-of-opportunity (ToO) observations during the mission with approximately one day lag in the ability to re-orient to the new target with minimal additional Delta-V ($\Delta v$) propulsion cost.

\begin{figure}
\centering
\vspace{3mm}
\includegraphics[width=0.95\columnwidth]{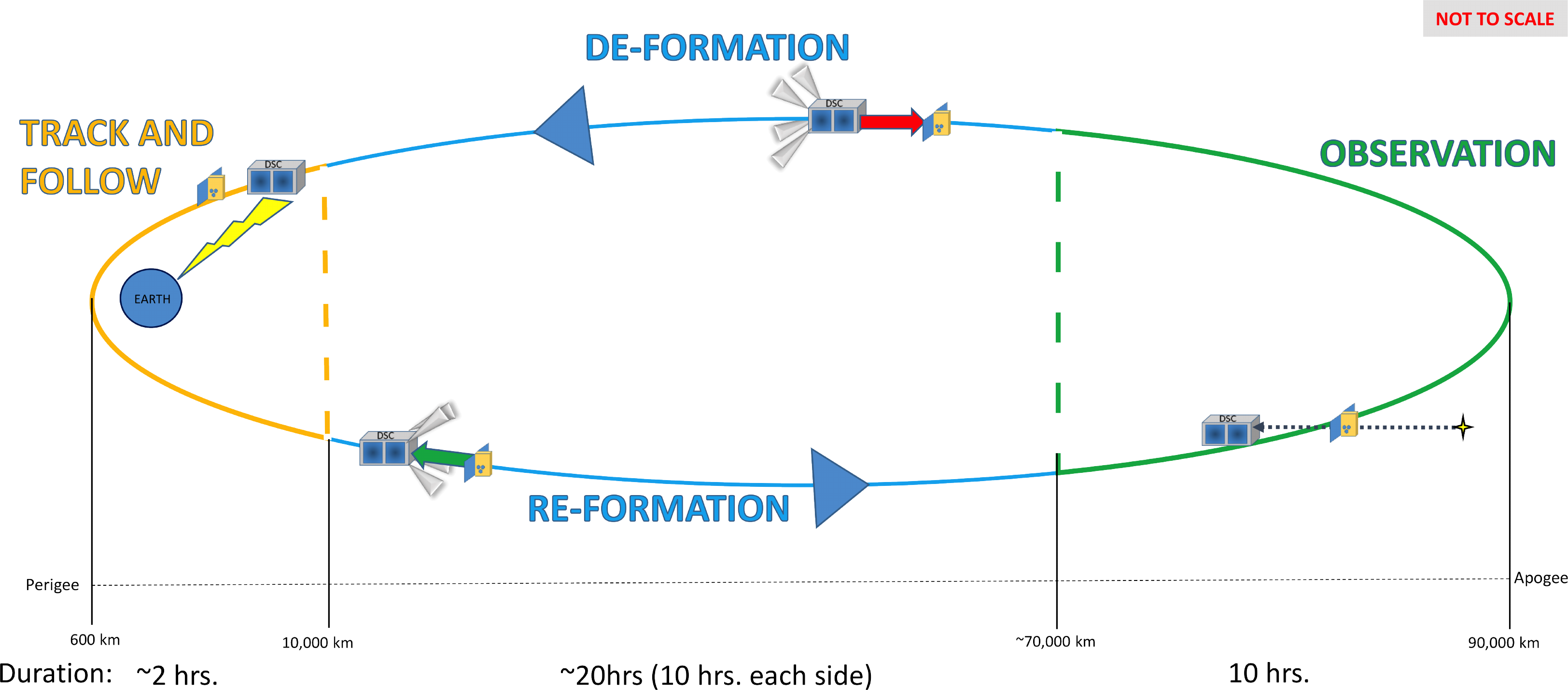}
\caption{VTXO concept of operations for one orbit. From Ref. \citenum{2020arXiv200612174K}.}
\label{figConOps}
\end{figure}

Figure \ref{figConOps} shows the concept of operations for a single VTXO orbit.  Near perigee, the two spacecraft fly in a loose formation with a 20 m separation in a desired inertial frame with good GPS attitude determination. Approximately 1 hour after perigee, a series of propulsion burns increase the separation and adjusts the inertial frame to the desired astrophysics target such that formation has 1 km separation and is in the desired precise formation 5 hours before apogee.  In precision formation flying science mode, the VTXO guidance, navigation, and control (GN\&C) system holds the two SmallSats with a 1 km $\pm < 1$ m separation with $\pm$ 5 mm transverse control and $< 1$ mm transverse position knowledge.  Combined with the 42 mas point-spread-function (PSF) of the VTXO telescope leads to a 55 mas (FWHM) resolution on VTXO's imaging. The science formation is held on a single target for 10 hours, that is there are no plans to change targets during the a single 10 hour science observation. At 5 hours after apogee, the precise formation is relaxed but still in an inertial frame, with the spacecraft separation adjusted to the target 20 m separation near perigee. At an altitude of $\sim$ 10,000 km, the DetectorSat adjusts its attitude to align the ground-link antenna to downlink the $\sim$ 200 Mbits of data and uplink any commands in an estimated 15 minute window needed to download the data.  After passing thru perigee, the process repeats for another orbit to enter another science formation-flying mode.

\acknowledgments 
 
This work was supported under NASA research announcement, NNH18ZDA001N-AS3 via proposal 18-AS318-0027 and NASA grant NNX09AG45G at NASA/GSFC, grant 80NSSC19K0123 at University of Maryland, Baltimore County (UMBC), and grant 80NSSC19K0695 at New Mexico State University (NMSU) and under NASA Cooperative Agreement Notice NNH15ZHA003C at NMSU by grant NM-NNX15AM73A.

\bibliography{spie2020vtxo} 

\begin{thebibliography}{10}

\bibitem{Giacconi1980}
{Giacconi}, R., ``{The Einstein X-ray Observatory},'' {\em Scientific
  American}~{\bf 242},  80--85 (Feb. 1980).

\bibitem{Swartz2010}
{Swartz}, D.~A., {Wolk}, S.~J., and {Fruscione}, A., ``{Chandra's First Decade
  of Discovery},'' {\em Proceedings of the National Academy of Science}~{\bf
  107},  7127--7134 (Apr. 2010).

\bibitem{2019JATIS...5b1001G}
{Gaskin}, J.~A., {Swartz}, D.~A., {Vikhlinin}, A., {{\"O}zel}, F., {Gelmis},
  K.~E., {Arenberg}, J.~W., {Bandler}, S.~R., {Bautz}, M.~W., {Civitani},
  M.~M., {Dominguez}, A., {Eckart}, M.~E., {Falcone}, A.~D.,
  {Figueroa-Feliciano}, E., {Freeman}, M.~D., {G{\"u}nther}, H.~M., {Havey},
  K.~A., {Heilmann}, R.~K., {Kilaru}, K., {Kraft}, R.~P., {McCarley}, K.~S.,
  {McEntaffer}, R.~L., {Pareschi}, G., {Purcell}, W., {Reid}, P.~B.,
  {Schattenburg}, M.~L., {Schwartz}, D.~A., {Schwartz}, E.~D., {Tananbaum},
  H.~D., {Tremblay}, G.~R., {Zhang}, W.~W., and {Zuhone}, J.~A., ``{Lynx X-Ray
  Observatory: an overview},'' {\em Journal of Astronomical Telescopes,
  Instruments, and Systems}~{\bf 5},  021001 (Apr. 2019).

\bibitem{2020arXiv200812810K}
{Krizmanic}, J., {Skinner}, G., {Arzoumanian}, Z., {Badilita}, V., {Gehrels},
  N., {Gendreau}, K., {Ghodssi}, R., {Gorius}, N., {Morgan}, B., {Mosher}, L.,
  and {Streitmatter}, R., ``{Phase Fresnel Lens Development for X-ray and
  Gamma-ray Astronomy},'' {\em arXiv e-prints} ,  arXiv:2008.12810 (Aug. 2020).

\bibitem{2020arXiv200612174K}
{Krizmanic}, J., {Shah}, N., {Harding}, A., {Calhoun}, P., {Purves}, L.,
  {Webster}, C., {Stochaj}, S., {Rankin}, K., {Smith}, D., {Park}, H.,
  {Boucheron}, L., {Kota}, K., {Corcoran}, M., {Shrader}, C., and {Naseri}, A.,
  ``{VTXO: the Virtual Telescope for X-ray Observations},'' {\em arXiv
  e-prints} ,  arXiv:2006.12174 (June 2020).

\bibitem{Skinner2001}
{Skinner}, G.~K., ``{Diffractive/refractive optics for high energy astronomy.
  I. Gamma-ray phase Fresnel lenses},'' {\em Astronomy and Astrophysics}~{\bf
  375},  691--700 (Aug. 2001).

\bibitem{Skinner2002}
{Skinner}, G.~K., ``{Diffractive-refractive optics for high energy astronomy.
  II. Variations on the theme},'' {\em Astronomy and Astrophysics}~{\bf 383},
  352--359 (Jan. 2002).

\bibitem{Krizmanic2005b}
{Krizmanic}, J., {Skinner}, G., and {Gehrels}, N., ``{Formation Flying for a
  Fresnel Lens Observatory Mission},'' {\em Experimental Astronomy}~{\bf 20},
  497--503 (Dec. 2005).

\bibitem{Morgan2004}
{Morgan}, B., {Waits}, C.~M., {Krizmanic}, J., and {Ghodssi}, R.,
  ``{Development of a deep silicon phase Fresnel lens using Gray-scale
  lithography and deep reactive ion etching},'' {\em Jour
  Microelectromechanical Systems}~{\bf 13},  113--230 (Feb. 2004).

\bibitem{Krizmanic2009}
{Krizmanic}, J. et~al., ``{Phase Fresnel Lens Development for X-ray and
  Gamma-ray Astronomy},'' in [{\em Proceedings of the 31st ICRC
  ({\L}\'{o}d\'{z})}{\nolinebreak\hspace{0.1em}]},  (2009).
\newblock Paper 1428.

\bibitem{Dennis2012}
{Dennis}, B.~R., {Skinner}, G.~K., {Li}, M.~J., and {Shih}, A.~Y., ``{Very
  High-Resolution Solar X-Ray Imaging Using Diffractive Optics},'' {\em Solar
  Physics}~{\bf 279},  573--588 (Aug. 2012).

\bibitem{NISTEX}
\lowercase{h}ttps://www.nasa.gov/mission\_pages/station/research/experiments/explorer/
  \newline Investigation.html/?\#id=7922.

\bibitem{2020arXiv200709287R}
{Rankin}, K., {Shah}, N., {Krizmanic}, J., {Stochaj}, S., and {Naseri}, A.,
  ``{Formation Flying Techniques for the Virtual Telescope for X-Ray
  Observations},'' {\em arXiv e-prints} ,  arXiv:2007.09287 (July 2020).

\bibitem{2020arXiv200709289R}
{Rankin}, K., {Krizmanic}, J., {Shah}, N., {Stochaj}, S., and {Naseri}, A.,
  ``{Virtual Telescope for X-Ray Observations},'' {\em arXiv e-prints} ,
  arXiv:2007.09289 (July 2020).

\end{thebibliography}
\bibliographystyle{spiebib} 

\end{document}